\documentclass[11pt]{article}
\usepackage{moriond,epsfig}

\def\Journal#1#2#3#4{{#1} {\bf #2}, #3 (#4)}

\def\NPB{{\em Nucl. Phys.} B}
\def\PLB{{\em Phys. Lett.}  B}
\def\PRL{\em Phys. Rev. Lett.}
\def\PRD{{\em Phys. Rev.} D}

\def\AP{{\em Ann. Phys.}}
\def\PTP{\em Prog. Theor. Phys.}
\def\JHEP{\em JHEP}

\def\be{\begin{equation}}
\def\ee{\end{equation}}
\def\bea{\begin{eqnarray}}
\def\eea{\end{eqnarray}}

\begin{document}
\vspace*{1cm}
\begin{flushright}
{DFPD 02/TH/10}
\end{flushright}
\vspace*{2cm}
\title{GAUGE SYMMETRY BREAKING ON ORBIFOLDS}

\author{ CARLA BIGGIO }

\address{Department of Physics, University of Padova and 
         I.N.F.N., Sezione di Padova, via Marzolo 8\\
          I-35131 Padova, Italy}

\maketitle\abstracts{ We discuss a new method for gauge symmetry
breaking in theories with one extra dimension compactified on the
orbifold $S^{1}/Z_{2}$.  If we assume that fields and their
derivatives can jump at the orbifold fixed points, we can implement a
generalized Scherk-Schwarz mechanism that breaks the gauge symmetry.
We show that our model with discontinuous fields is equivalent to
another with continuous but non periodic fields; in our scheme
localized  lagrangian terms for bulk fields appear.  
}


Symmetries (and their breaking) play a fundamental role in theoretical
high energy physics.  In theories with extra dimensions new methods
for symmetry breaking appear, such as the Scherk-Schwarz
mechanism~\cite{Scherk-Schwarz}, the Hosotani
mechanism~\cite{Hosotani} and the orbifold projection~\cite{orbifold}.

In this paper we will focus on theories with one extra dimension
compactified on the orbifold $S^{1}/Z_{2}$ where the radius of the
circle $R$ is of order of $1/M_{GUT}$~\cite{Fayet}, with
$M_{GUT}\approx 10^{16}Gev$, and the orbifold $Z_{2}$ is constructed
by identifying points $(x,y)$ with points $(x,-y)$, where $x$ and $y$
are respectively the coordinates of the four and the fifth
dimensions. The orbifold projection introduces two fixed points, at
$y=0$ and $y=\pi R$. While on the circle the only boundary condition
we can assign to fields is the one corresponding to the translation of
$2\pi R$ (twist condition), on the orbifold we have to fix the parity
and the behavior of fields at the fixed points, that means new
possibilities are allowed~\cite{BFZ}.

If we consider a set of $n$ 5-dimensional bosonic fields $\phi
(x,y)$~\cite{our}, the self-adjointness of the kinetic operator
requires that we assign conditions also on the $y$-derivative of $\phi
(x,y)$. We look for boundary conditions in the following class:\\
\begin{equation}
\label{generalbc}
\left(
\begin{array}{c}
\phi\\
\partial_y\phi
\end{array}
\right)(\gamma^+) =
V_\gamma~\left(
\begin{array}{c}
\phi\\
\partial_y\phi
\end{array}
\right)(\gamma^-)~~,\\
\end{equation}
where $\gamma=(0, \pi,\beta)$, $\gamma^\pm=(0^\pm,\pi R^\pm, y^\pm)$,
and $V_\gamma$ are constant $2n$ by $2n$ matrices.  We have defined
$0^\pm\equiv\pm\xi$, $\pi R^\pm\equiv\pi R\pm\xi$, with $\xi$ small
positive parameter, $y^-\equiv y_0$ and $y^+\equiv y_0+2 \pi R$, with
$y_0$ for convenience chosen between $-\pi R+\xi$ and $-\xi$.

We now constrain the matrices $V_\gamma$. We require, first, that the
boundary conditions preserve the self-adjointness of the kinetic
operator $-\partial_{y}^{2}$ and, second, that physical quantities
remain periodic and continuous even if fields are not.  If the theory
is invariant under the global transformations of a group G, these
requirements imply the following form for the $V_\gamma$ matrices:
\begin{equation}
\label{generalVgamma}
V_\gamma=
\left(
\begin{array}{cc}
U_\gamma & 0        \\
0        & U_\gamma
\end{array}
\right)~,
\end{equation}
where $U_\gamma$ is an orthogonal $n$-dimensional representation of G.
Finally we should take into account consistency conditions among the
twist, the jumps and the action of orbifold on the fields, defined by
the following equation:
\begin{equation}
\label{z_on_phi}
\phi (-y) = Z \phi (y)~~.
\end{equation}
We have:
\begin{eqnarray}
\label{consistency}
U_\gamma~Z~ U_\gamma &=& Z \qquad\qquad \gamma\in (0, \pi , \beta) \nonumber\\
\left[U_0,U_\beta\right]&=& 0\\
\left[U_\pi,U_\beta\right]&=& 0~~.\nonumber
\end{eqnarray}
If $\left[Z,U_\gamma\right] = 0$ and 
$\left[U_0,U_\pi\right] = 0$
there is a basis in which $U_\gamma$ are diagonal with elements $\pm
1$; in this case twist and jumps are defined by discrete parameters.
However if $\left[Z,U_\gamma\right] \ne 0$ or $\left[U_0,U_\pi\right]
\ne 0$ continuous parameter can appear in $U_\gamma$ and then in
eigenfunctions and eigenvalues.\\

To show how we can exploit these boundary conditions to break a
symmetry, we focus on the simple case of one real scalar field.  We
start by writing the equation of motion for $\phi$
\begin{equation}
\label{eqofmotion}
-\partial^2_y \phi = m^2 \phi~~,
\end{equation}
in each region $y_q<y<y_{q+1}$, where $y_q\equiv q \pi R$ and $q\in
Z$. We have defined the mass $m$ through the 4-dimensional equation
$\partial^2 \phi= m^2 \phi$.  The solutions of these equations can be
glued by exploiting the boundary conditions $V_0$ and $V_\pi$, imposed
at $y=y_{2q}$ and $y=y_{2q+1}$, respectively. Finally, the spectrum
and the eigenfunctions are obtained by requiring that the solutions
have the twist described by $V_\beta$.  In this theory the group of
global symmetry is a $Z_2$, so that $V_\gamma$ can be $\pm 1$.

We focus on one example in which we compare the usual Scherk-Schwarz
mechanism with our generalized one. We consider even fields and the
following two sets of boundary conditions:
\begin{equation}
\label{example}
(U_0, U_\pi, U_\beta)~  = ~
\left\{
\begin{array}{cc}
(+1, +1, -1) & ~~~~~(\textrm{A}) \\
(+1, -1, +1) & ~~~~~(\textrm{B}) 
\end{array}
\right. .
\end{equation}
In the first case fields are continuous but antiperiodic; the solution
to the eq. of motion with these boundary conditions is:
\begin{equation}
\label{solutionA}
\phi^A (x,y) = \alpha (x) ~ cos(my) \qquad \qquad  \qquad \quad m R = n + \frac{1}{2} ~~.
\end{equation}
In the case (B) fields are periodic but with a jump in $y=\pi R$; the solution is:
\begin{equation}
\label{solutionB}
\phi^B (x,y) = \beta (x)~\epsilon \left (\frac{y}{2} +\frac{\pi R}{2} \right )
~cos(my)~~~~~~ m R = n + \frac{1}{2} ~~,
\end{equation}
where $\epsilon (y)$ is the sign function on $S^1$.
\begin{figure}[!th]
\centerline{
\epsfig{figure=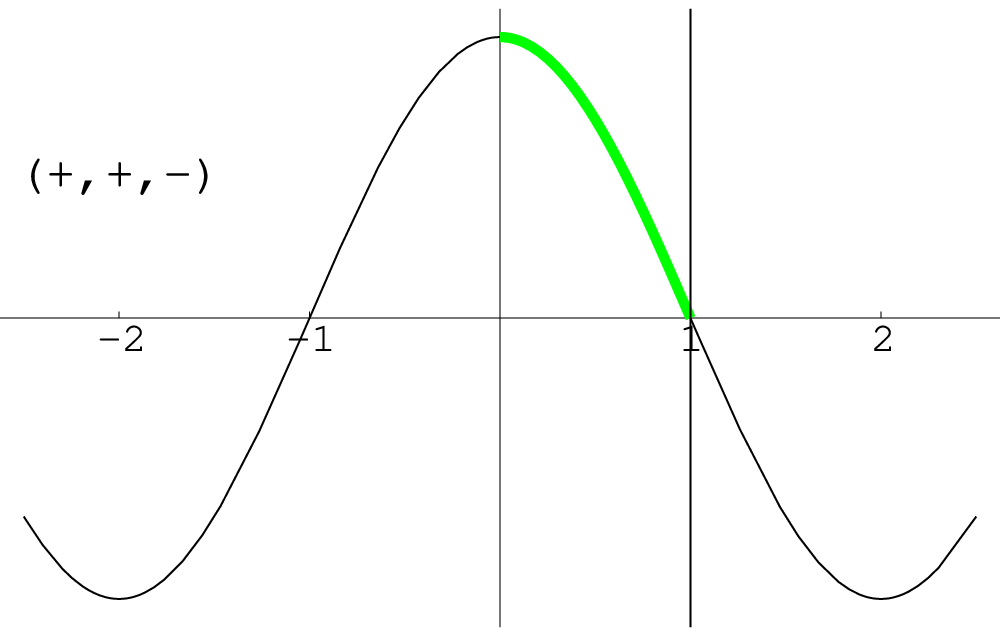,height=4cm,width=7cm}~~~
\epsfig{figure=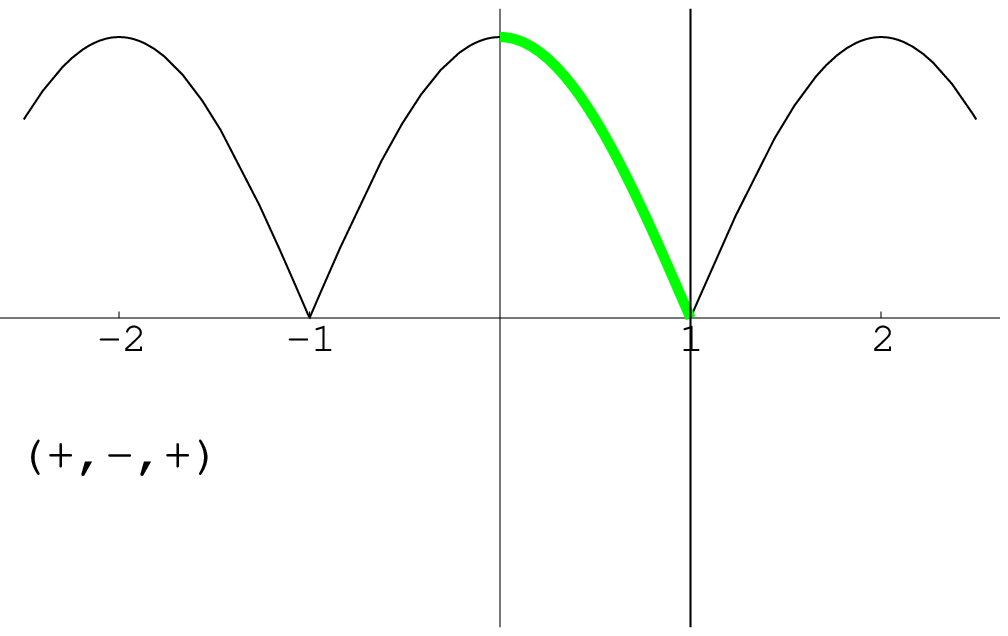,height=4cm,width=7cm}}
\caption{Eigenfunctions of $-\partial_y^2$ versus $y/(\pi R)$, in cases (A) and (B).}
\label{graphicsAB}
\end{figure}
The spectrum is the same in both cases and it is the Kaluza-Klein
tower $n/R$ shifted by $1/2R$. This is just what we expected in case
(A), because this is precisely the Scherk-Schwarz mechanism: when
fields are not periodic, their spectra are shifted from the
Kaluza-Klein levels by an amount depending on the twist and zero modes
are removed.  If $\phi$ is a gauge field of a multiplet containing
also untwisted fields, this is a mechanism for symmetry breaking.  As
we have already observed the (B)-spectrum is identical to the
(A)-spectrum; this means that we can break a symmetry not only through
a twist, but also with discontinuities.  While the spectra are
identical, eigenfunctions (see fig. \ref{graphicsAB}) are different,
but they are related by a simple field redefinition:
\begin{equation}
\label{fieldredef}
\phi^B (x,y) ~ = ~ \epsilon \left (\frac{y}{2} +\frac{\pi R}{2}\right ) ~ \phi^A (x,y)~~.
\end{equation}

The equation (\ref{fieldredef}) is a map between a system in which
fields are continuous and twisted (A) and another in which fields are
discontinuous and periodic (B).  In order to show the two schemes are
completely equivalent, we should perform the field redefinition at the
level of the action and, using the eq. of motion, recover the
eigenfunction and the spectrum. After the field redefinition, the
lagrangian for the massless real scalar field becomes:
\begin{equation}
\label{scalar_lagrangian}
\mathcal{L}(\phi,\partial\phi) =
\frac{1}{2} \epsilon^{2} \partial_{M}\phi~\partial^{M}\phi  
- 2~\epsilon~\delta_{\pi R}~ \phi~\partial_{y}\phi + 
 2~\delta_{\pi R}^{2}~ \phi^{2}~~,
\end{equation}
where $\epsilon = \epsilon (y/2 + \pi R/2)$ and $\delta_{y_{0}} =
\delta (y-y_{0})$. The new lagrangian contains two quadratic terms
localized at $y=\pi R$. Now we want to derive the equation of motion
for $\phi$, but the naive application of the variational principle
would lead to inconsistent results: indeed this works with continuous
functions, but $\phi$ are discontinuous. In order to derive the
correct equation of motion we regularize $\epsilon$ with a smooth
function $\epsilon_{\lambda}$ wich reproduces $\epsilon$ in the limit
$\lambda \rightarrow 0$. If we now rewrite the lagrangian
(\ref{scalar_lagrangian}) in term of $\epsilon_{\lambda}$ and
continuous fields, derive the equation of motion and then perform the
limit $\lambda \rightarrow 0$, we obtain:
\begin{equation}
\label{eq.ofm._scalar}
 \epsilon \partial_{y}^{2}\phi 
 - 4 \delta_{\pi R} \partial_{y}\phi 
 - 2 \delta_{\pi R}' \phi +
 \epsilon m^{2} \phi = 0
\end{equation}
Away from the point $y=\pi R$ this equation reduces to the one for
continuous fields, while integrating it around the fixed point the
singular terms give us the expected jumps. With this procedure we have
shown that the continuous framework is really equivalent to the
discontinuous one.

If we consider two or more real scalar fields such as a complex scalar
field or a gauge field, continuous parameters are allowed in the
boundary conditions and then they are also present in the
eigenfunctions and in the spectra. It is interesting to note that,
being the shift from the Kaluza-Klein spectrum continuous, we can go
continuously from a phase in which the symmetry is broken to another
in which it is unbroken.\\

\begin{table}[!t]
\caption{Boundary conditions, eigenfunctions and spectra for fields in
the Kawamura model in the traditional scheme ($2^{nd}$, $3^{rd}$ and
$6^{th}$ columns) and in our framework ($4^{th}$, $5^{th}$ and
$6^{th}$ columns).}
\vspace{0.4cm}
\centering{
\begin{tabular}{|c||c|c||c|c||c|}
\hline
& & & & & \\
Fields & $(Z_2,U_\beta)$ & ${\textrm{Kawamura} \atop \textrm{eigenfunctions}}$
 & $(Z_2,U_\pi)$ & ${\textrm{Our} \atop \textrm{eigenfunctions}}$
 & ${\textrm{Spectrum} \atop m}$ \\
& & & & & \\
\hline
& & & & & \\
$A^{a}_{\mu}, \lambda_{L}^{2a}, H^{D}_{u}, H^{D}_{d}$ & 
$(+,+)$  & $cos(my)$ & $(+,+)$  & $cos(my)$ & $\frac{2n}{R}$ \\
& & & & & \\
\hline
& & & & & \\
$A^{\hat{a}}_{\mu}, \lambda_{L}^{2\hat{a}}, H^{T}_{u}, H^{T}_{d}$ & 
$(+,-)$  & $cos(my)$ & $(+,-)$  & $\epsilon (\frac{y}{2}+\frac{\pi R}{2})cos(my)$ & $\frac{2n+1}{R}$ \\
& & & & & \\
\hline
& & & & & \\
$A^{\hat{a}}_{5}, \Sigma^{\hat{a}},\lambda_{L}^{1\hat{a}}, \hat{H}^{T}_{u}, \hat{H}^{T}_{d}$ & 
$(-,-)$  & $sin(my)$ & $(-,-)$  & $\epsilon (\frac{y}{2}+\frac{\pi R}{2})sin(my)$ & $\frac{2n+1}{R}$ \\
& & & & & \\
\hline
& & & & & \\
$A^{a}_{5}, \Sigma^{a},\lambda_{L}^{1a}, \hat{H}^{D}_{u}, \hat{H}^{D}_{d}$ & 
$(-,+)$  & $sin(my)$ & $(-,+)$  & $sin(my)$ & $\frac{2n+2}{R}$ \\
& & & & & \\
\hline
\end{tabular}}
\label{table}
\end{table}

We use now our mechanism to break the symmetries of a grand unified
theory defined on the orbifold $S^1/Z_2$~\cite{models,Kawamura}. The
model we consider is the one constructed by Kawamura~\cite{Kawamura}
in which the lagrangian is invariant under the transformations of an
$SU(5)$ gauge group and $N=2$ (from the 4-dimensional point of view)
supersymmetry. The fields content is the following: there is a vector
multiplet $V = (A^\alpha_M, \lambda^{1\alpha}_L, \lambda^{2\alpha}_L,
\Sigma^\alpha )$, with $\alpha = 1,...,24$ and $M=\mu ,5$ which forms
an adjoint representation of $SU(5)$, and two hypermultiplets $H^1$
and $H^2$, equivalent to four chiral multiplets $H_5$,
$\hat{H_{\overline{5}}}$, $\hat{H_{5}}$ and $H_{\overline{5}}$ which
form two fundamental representations of $SU(5)$. These are multiplets
of $N=1$ supersymmetry in 5 dimensions, that is equivalent to $N=2$
supersymmetry in 4 dimensions.  The gauge vector bosons are the fields
$A^\alpha_M$: $A^a_M$ with $a=1,...,12$ are the vector bosons of the
standard model, while $A^{\hat{a}}_M$ with $\hat{a}=13,...,24$ are
those of the coset $SU(5)~/~SU(3)\times SU(2)\times U(1)$.

In table \ref{table} (second column) parity and twist assignments for
all fields are shown, together with the resulting eigenfunctions and
spectra (third and last columns).  We see that these parity assigments
alone break supersymmetry from $N=2$ to $N=1$, because zero modes
survive only for even fields (without twist the spectrum would be
$n/R$ in all cases). When we assign a non trivial twist to fields,
their spectra are shifted from the usual Kaluza-Klein tower. With the
assignments of table \ref{table} the only fields that mantain a zero
mode are the vector bosons of the standard model, $A_\mu^a$, their
supersymmetric partners, $\lambda^{2a}_L$, and two Higgs doublets
$H^{D}_{u}$ and $H^{D}_{d}$: precisely the fields of the Minimal
Supersymmetric Standard Model, that is $SU(3)_c\times SU(2)_L\times
U(1)_Y$ with $N=1$ supersymmetry.  We observe that with these boundary
conditions also the doublet-triplet splitting problem is naturally
solved, because the lightest triplets modes are of order of $1/R$,
while doublets have zero modes.

Up to now only gauge and Higgs fields and their supersymmetric
partners have been considered. If we introduce fermions and we assign
the appropriate boundary conditions~\cite{AF} we can see that fast
proton decay is avoided. Then this model can be a realistic Grand
Unified Theory and we adopt it to show in details how our mechanism
for gauge symmetry breaking works.

In our framework parity assignments are identical to the original
ones, but, instead of a twist, we require that some fields jump in
$y=\pi R$. Among gauge fields only the vector bosons of the coset
$SU(5)~/~SU(3)\times SU(2)\times U(1)$ jump. In the last three columns
of table \ref{table} boundary conditions, eigenfunctions and
eigenvalues are shown. We can observe that the spectra are the same of
the Kawamura model for all fields, while eigenfunctions are identical
for continuous fields but different for jumping fields. If we look at
their explicit forms we observe we are just in the case studied before
with one real scalar field. Performing the field redefinition of
eq. (\ref{fieldredef}), where `A' fields are Kawamura's eigenfunctions
and `B' fields are ours, we see the two frameworks are completely
equivalent.

What about the lagrangians? Following the lines of what we did in the
scalar case, we can perform the field redefinition at the level of the
action in order to find the localized terms related to jumps in this
theory. For simplicity we consider only the Yang-Mills term of the
lagrangian, neglecting both the supersymmetric part and the Higgs
terms:
\begin{equation}
\label{lagrangian_SU(5)}
\mathcal{L} = -\frac{1}{4} F^{\alpha}_{MN} F^{\alpha MN}~~.
\end{equation}
Applying the redefinition (\ref{fieldredef}) this becomes:
\begin{eqnarray}
\label{last-lagrangian}
& -\frac{1}{4} \tilde{F}^{a}_{MN} \tilde{F}^{a MN}
+\frac{1}{2} \epsilon^{2} f^{a\hat{b}\hat{c}} A^{\hat{b}}_{M} A^{\hat{c}}_{N}
   \tilde{F}^{a MN} + & \nonumber \\[0.15cm] 
& -\frac{1}{4} \epsilon^{4} f^{a\hat{b}\hat{c}} f^{a\hat{d}\hat{e}} 
   A^{\hat{b}}_{M} A^{\hat{c}}_{N} A^{\hat{d}M} A^{\hat{e}N} 
-\frac{1}{4} \epsilon^{2} F^{\hat{a}}_{MN} F^{\hat{a}MN} + & \\[0.25cm]
& -2~\delta_{\pi R}^{2}~ 
   (A^{\hat{a}}_{N} A^{\hat{a}N} - A^{\hat{a}}_{5} A^{\hat{a}5}) 
+2~\epsilon~\delta_{\pi R}~ F^{\hat{a}}_{5N} A^{\hat{a}N}~~, & \nonumber 
\end{eqnarray} 
where $\tilde{F}^{a}_{MN} = \partial_{M} A^{a}_{N} -\partial_{N}
A^{a}_{M} - f^{abc} A^{b}_{M} A^{c}_{N}$.  We observe that the last
two terms are localized at the fixed point $y=\pi R$. While the first
is simply a bilinear term, as the ones found in the case of one scalar
field, the last contains also a trilinear part:
\begin{equation}
\label{brane-interaction}
2~\epsilon ~\delta_{\pi R} ~F^{\hat{a}}_{5N} A^{\hat{a}N} =
2~\epsilon ~\delta_{\pi R}~ 
  [ (\partial_{5} A^{\hat{a}}_{N}-\partial_{N} A^{\hat{a}}_{5}) A^{\hat{a}N}
   + f^{\hat{a} b\hat{c}} A^{\hat{a}N} A^{b}_{N} A^{\hat{c}}_{5} ].
\end{equation} 
This last term is proportional to the fields $A^{\hat{c}}_{5}$ that
are just the `would-be' Goldstone bosons that give mass to the
Kaluza-Klein modes of the gauge vector bosons of the coset $SU(5) /
SU(3)\times SU(2)\times U(1)$.  

In the previuos pages we always started from the continuous theory and
then we derived the localized lagrangian terms, exploiting a relation
found between smooth and jumping eigenfunctions. But our reasoning can
be reversed: we can try to reabsorb localized terms for bulk fields
through a field redefinition. This kind of terms are very common in
theories with extra dimensions~\cite{brane-terms}, and showing they are
only the effect of some discontinuities of the fields would improve
our control of the theory. In our work we have shown that in some cases
this is possible.\\

In summary in this paper we have shown how the Scherk-Schwarz
mechanism works with discontinuous fields, in particular with bosonic
fields. We have analyzed in detail the constraints on the generalized
boundary conditions and we have shown the connection between the
smooth framework and the discontinuous one in the case of one real
scalar field. Then we have applied our mechanism to a Grand Unified
Theory based on the $SU(5)$ gauge group that breaks down to
$SU(3)\times SU(2)\times U(1)$ and we have shown how localized terms
for bulk fields appear in the lagrangian. We have observed that these
terms are strictly related to discontinuities of fields at the fixed
points and that in some cases they can be reabsorbed through a field
redefinition.


\section*{Acknowledgments}
I would like to thank Ferruccio Feruglio for the enjoyable and
fruitful collaboration on which this talk is based and Fabio Zwirner
for useful discussions. Many thanks go also to the organizers of the
`Rencontres de Moriond' for the pleasant and relaxed atmosphere of the
conference. This work was partially supported by the European Program
HPRN-CT-2000-00148 (network `Across the Energy Frontier') and by the
European Program HPRN-CT-2000-00149 (network `Physics at Colliders').


\end{document}